\documentclass[aps,prb,showpacs,twocolumn,superscriptaddress,floatfix]{revtex4-1}
\usepackage[colorlinks,bookmarks=false,citecolor=blue,linkcolor=red,urlcolor=blue]{hyperref}
\usepackage{epsfig}
\usepackage{tikz}
\usepackage{graphicx,comment}

\usepackage{dcolumn}
\usepackage{bm}

\usepackage{amsmath,amssymb}

\usepackage{bbold}

\newcommand{\bea}{\begin{eqnarray}}          
\newcommand{\eea}{\end{eqnarray}}          
\newcommand{\la}{\langle}          
\newcommand{\ra}{\rangle}

\allowdisplaybreaks

\begin{document}
\title{Ice on curved surfaces: defect rings and differential local dynamics}
\author{Adhitya Sivaramakrishnan}
\email{adhityasubramaniam@gmail.com}
\affiliation{Department of Applied Physics, KTH Royal Institute of Technology, SE-106 91 Stockholm, Sweden}
\author{R. Ganesh}
\email{r.ganesh@brocku.ca}
\affiliation{Department of Physics, Brock University, St. Catharines, Ontario L2S 3A1, Canada}

\date{\today}

\begin{abstract}
Ice systems are prototypes of locally constrained dynamics. This is exemplified in Coulomb-liquid phases where a large space of configurations is sampled, each satisfying local ice rules. Dynamics proceeds through `flipping' rings, i.e., through reversing arrows running along the edges of a polygon. We examine the role of defect rings in such phases, with square-ice as a testing ground. When placed on a curved surface, the underlying square lattice will form defects such as triangles or pentagons. 
We show that triangular defects are statistically more `flippable' than the background. 
In contrast, pentagons and larger polygons are less flippable.
In fact, flippability decreases monotonically with ring size, as seen from a Pauling-like argument. 
As an explicit demonstration, we wrap the square ice model on a sphere. 
We start from an octahedron and perform repeated rectifications, producing a series of clusters with sphere-like geometry. They contain a fixed number of defect triangles in an otherwise square lattice.   
We numerically enumerate all ice-rule-satisfying configurations. Indeed, triangles are flippable in a larger fraction of configurations than quadrilaterals. The obtained flippabilities are in broad agreement with the Pauling-like estimates. 
As a minimal model for dynamics, we construct a Hamiltonian with quantum tunnelling terms that flip rings. The resulting ground state is a superposition of all ice configurations. The dominant contribution to its energy comes from localized resonance within triangles. 
Our results suggest local dynamics as a promising observable for experiments in spin ice and artificial ice systems. They also point to hierarchical dynamics in materials such as ice V that contain rings of multiple sizes.  
\end{abstract}

\maketitle

\section{Introduction} 
Ice, as formed on earth, can be viewed as half-solid: its oxygen atoms are at fixed positions while hydrogen atoms are free to move. Their motion is, however, heavily constrained by local `ice rules'\cite{BernalFowler1933}. Each oxygen atom must have two hydrogen atoms that are near and two that are farther, forming a two-in-two-out configuration.  
The same physics appears in spin ice materials, where spins on a tetrahedron must satisfy a two-in-two-out rule\cite{BramwellGingras2001,Castelnovo2012,Bramwell_2020,Udagawa2021}. Analogues have been found in soft matter\cite{TiernoRMP2019}, artificial spin ice\cite{NisoliRMP2013} and Rydberg atom ensembles\cite{Glaetzle2014}. In any ice system, the local ice rules are satisfied by a large ensemble of configurations. When the system samples this ensemble ergodically, it is said to be in a Coulomb liquid phase\cite{Henley2010}. 
Such phases have been extensively studied in homogeneous ice systems. In this article, we examine the role of inhomogeneities in the form of defect rings. We demonstrate that they may locally enhance or suppress dynamics. For example, near a defect ring in water ice, hydrogen atoms may move more frequently as compared to defect-free regions.

Enumeration of ice configurations is a problem of great interest in statistical physics\cite{Caravelli2021}. An early, influential contribution was Pauling's estimate for the entropy, based on local configurations around a single site\cite{Pauling1935}. A landmark test case appears in square ice\cite{Lieb1967}, where the entropy can be obtained rigorously in the limit of large system sizes. Pauling's estimate captures the correct functional form for the dependence on system size, with a small discrepancy in the numerical value of the exponent. 
In the following sections, we examine local contributions to entropy using a Pauling-like approach. We enumerate configurations within a ring, based solely on local ice-rules. To test our ideas, we adapt square ice 
to spherical geometries, with defects in the form of triangles.

To have a two-in-two-out ice rule, we require a background lattice that has coordination number four. This is naturally realized in the square lattice in two dimensions and in the diamond lattice in three dimensions. Many other lattices can be found in the phase diagram of water ice, where each oxygen atom is surrounded by four others\cite{Chaplin,Petrenko1999}. For example, terrestrial ice is in the Ih phase with oxygens forming a Wurtzite lattice. The shortest rings here have six oxygen atoms that do not lie on a single plane. The motion of hydrogen ions (protons) occurs predominantly within these rings\cite{Bove2009,Marx2014,Yen2015}. This can be contrasted with ice V, a crystalline phase that occurs at around 253 K and 500 MPa\cite{Kamb1967,Petrenko1999}. Its structure contains 4-membered, 5-membered, 6-membered, and 8-membered rings. A similar situation arises at stacking faults\cite{Pirzadeh2010} and in amorphous ice\cite{Salzmann2015} with rings of multiple sizes. In such systems, we may expect rings to have varying levels of dynamics, e.g., with the smallest rings having more mobile protons. 
Below, we present arguments that support this conjecture.

\section{A ring of size M: configurations and flippability}
\label{sec.polygon}
We enumerate local configurations of a ring within a generic ice system. In the spirit of Pauling's estimate, the only constraint we impose is the ice rule at each of the ring's vertices. The environment places no restrictions, e.g., long-range correlations arising from rings of the environment are ignored. In later sections, we examine the accuracy of these arguments using a series of clusters that host rings of two sizes.

\begin{figure}
\includegraphics[width=3in]{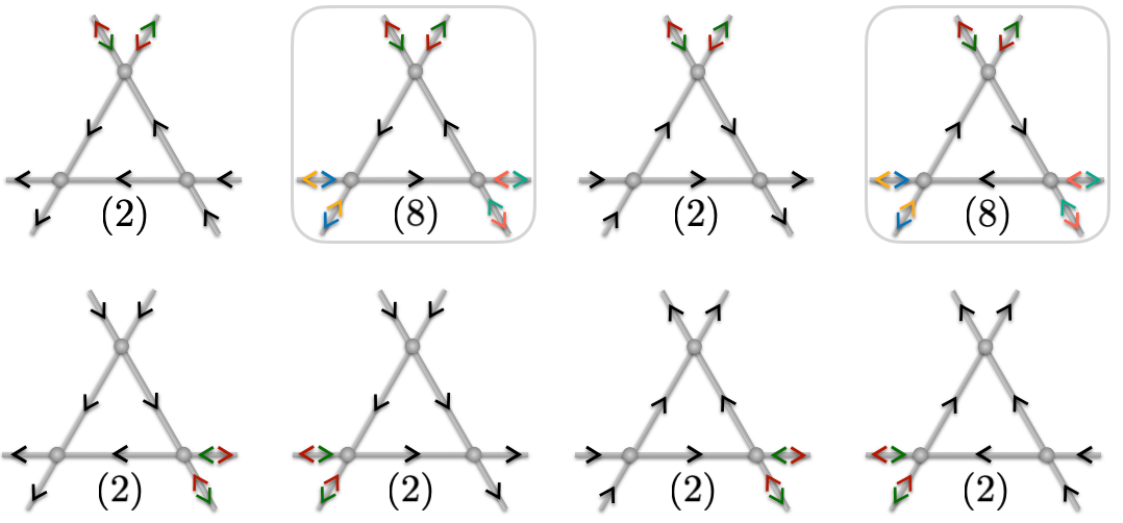}
\caption{Configurations on a triangle ring. All possible configurations of the internal arrows are shown. In some cases, ice rules at a vertex allow for two configurations of external arrows: one pointing inwards, the other outwards. These bonds are shown with two arrowheads. The degeneracy of each configuration, after accounting for choices for external arrows, is shown in parentheses. Configurations where the triangle is flippable are shown enclosed in a box.
  }
\label{fig.triconfigs}
\end{figure}

\begin{figure}
\includegraphics[width=3in]{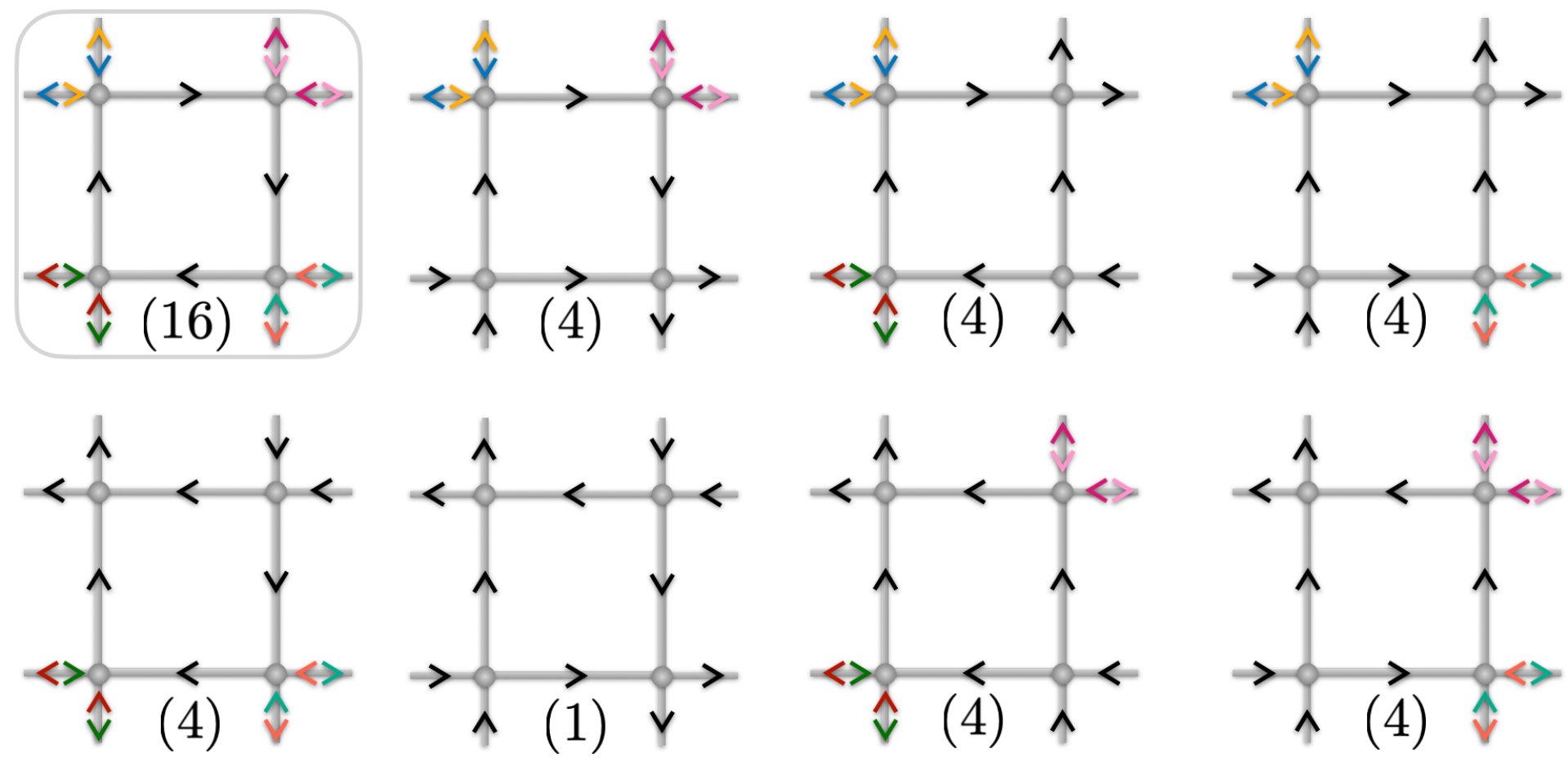}\\
\includegraphics[width=3in]{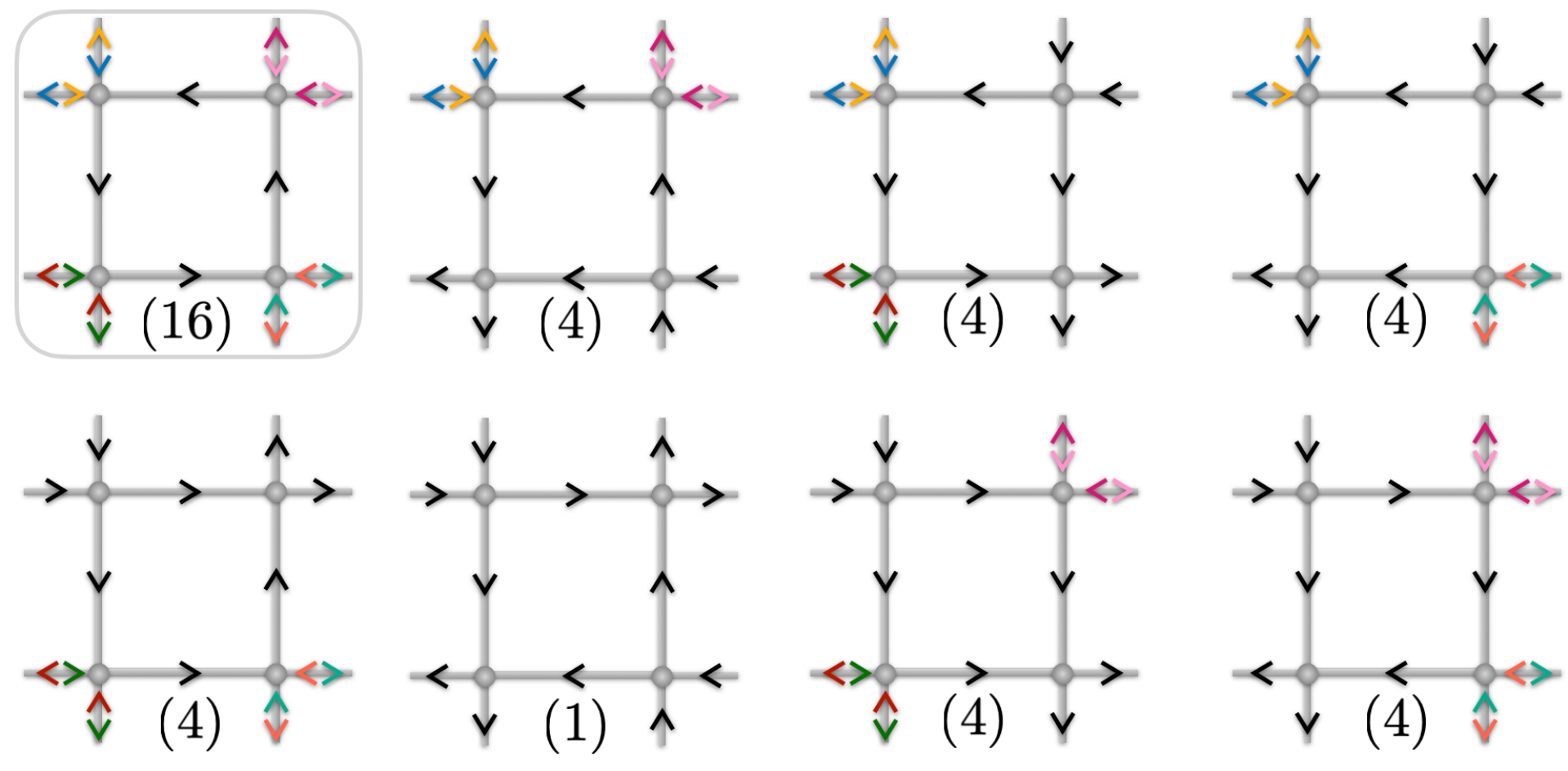}
\caption{Configurations on a quadrilateral ring, with all possible internal-arrow configurations. External arrows with double arrowheads indicate a two-fold choice at the corresponding vertex: that of choosing one external arrow to point inward while the other points outward. These choices lead to the degeneracy shown in parentheses. Flippable configurations are shown enclosed in a box.}
\label{fig.sqconfigs}
\end{figure}

Consider a polygon ring with $M$ vertices. Each vertex must have four bonds connected to it. We divide them into two pairs: internal and external. The former connect to neighbouring vertices within the polygon while the latter take us out of the ring. On each of these four bonds, we draw an arrow that may point inward (towards reference vertex) or outward (away from reference vertex). For the internal edges, an outgoing arrow on one vertex becomes an incoming arrow for a neighbouring vertex and vice versa. On the external edges, we place no constraints assuming that the surroundings allow for all sets of orientations with equal weight.  
We next enforce the ice rules: on each vertex, we must have two incoming and two outgoing arrows. We enumerate all such configurations. If two configurations differ in the orientation of external arrows, they are counted as being distinct. Relegating details to Appendix.~\ref{app.polyM}, the total number of allowed configurations is $1+ 3^M$. For example, a triangle ($M=3$) allows for 28 possibilities as shown in Fig.~\ref{fig.triconfigs}, while a quadrilateral ($M=4$) allows for 82 possibilities, as shown in Fig.~\ref{fig.sqconfigs}.

We next examine if the ring is flippable. Can the internal arrows be reoriented without changing any of the external arrows, while satisfying the ice rule at every vertex? This is only possible if all internal arrows have the same sense. That is, as we move along the polygon, all arrows must point along the direction of motion or all must point opposite. In such configurations, each vertex has one incoming internal arrow and one outgoing internal arrow. The external arrows must also be one-in, one-out. There is a two-fold choice at each vertex -- of choosing one of the external arrows to point outward. The total number of flippable configurations is $2 \times 2^M$. The first factor of $2$ comes from choosing all internal arrows to be along/opposite the direction of motion. The factor of $2^M$ comes from the two-fold choice of external arrows at each vertex. 

We now define `flippability', $f_M$, as the likelihood of the ring being flippable. Assuming all configurations to be equally likely, we find 
\begin{equation}
f_M = \frac{2^{M+1}}{1+3^M}.
\label{eq.fM}
\end{equation}

We illustrate two specific examples. Fig.~\ref{fig.triconfigs} depicts configurations of a triangle ($M=3$). We have 28 configurations in total, out of which 16 are flippable. This leads to a flippability of $16/28\sim 57.14\%$. Fig.~\ref{fig.sqconfigs} depicts configurations of a quadrilateral ring ($M=4$). Out of 82 configurations, 32 are flippable. The flippability comes out to be $32/82\sim 39.02\%$. More generally, from Eq.~\ref{eq.fM}, we see that $f_M$ is a decreasing function of $M$. Flippability decreases monotonically with ring-size. Triangles have the highest flippability, followed by quadrilaterals, pentagons and so on.

\section{Square ice on a sphere}

\begin{figure*}
\includegraphics[width=7in]{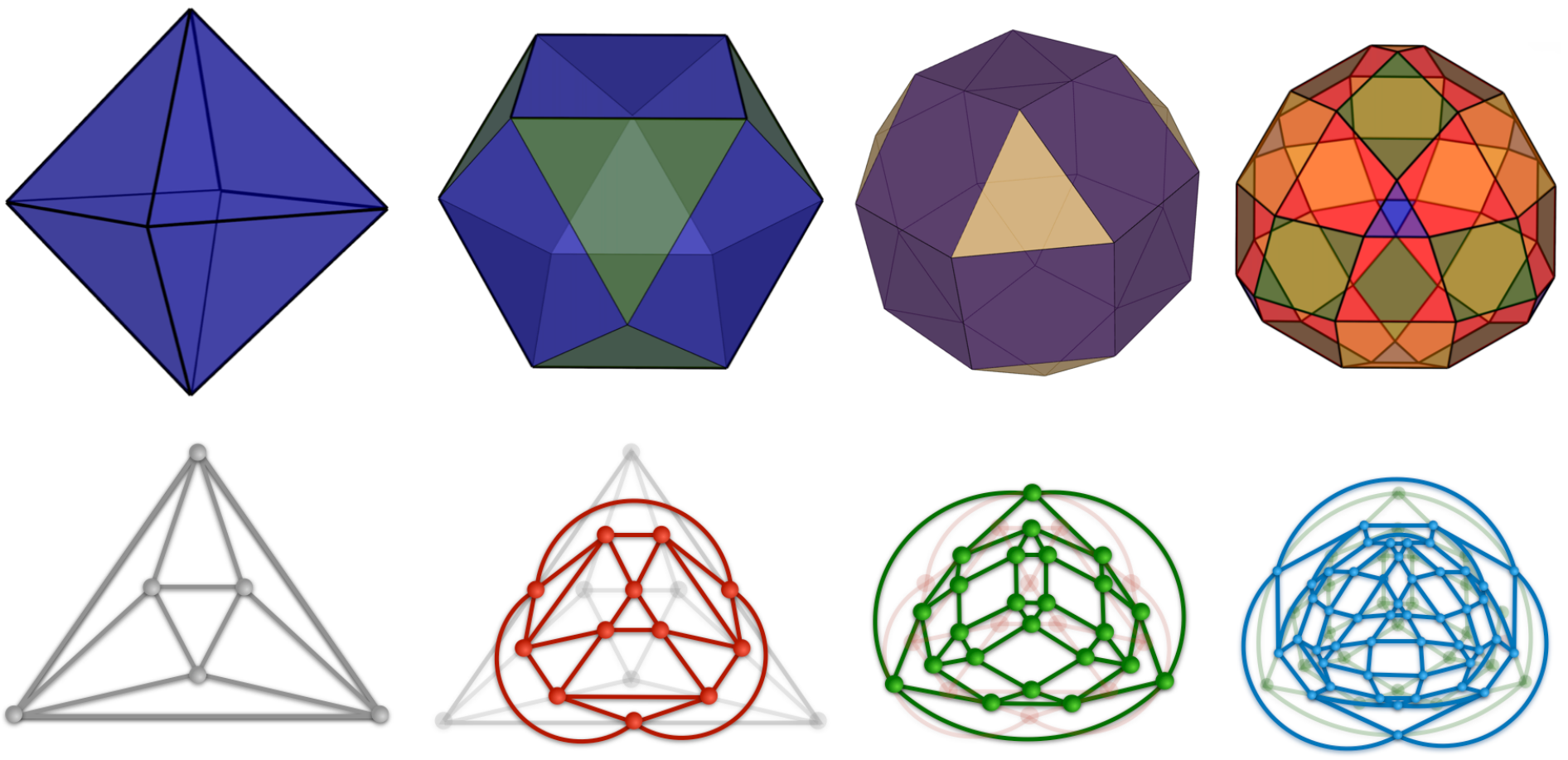}
\caption{The first four clusters of our series. The top row shows a three-dimensional view, while the bottom row shows a planar representation (Schlegel diagram). From left to right, we have an octahedron, a cuboctahedron, a rhombicuboctahedron and an expanded cuboctahedron. From left to right, each subsequent cluster is obtained by a rectification operation on the preceding cluster. Equivalently, each subsequent Schlegel diagram is obtained as a medial graph of its predecessor. To show this, each Schlegel diagram is shown against an outline of that of the preceding cluster.}
\label{fig.clusters}
\end{figure*}

The square lattice is typically studied with the topology of a torus, by enforcing periodic boundary conditions in two perpendicular directions. A square grid cannot be wrapped on a sphere as this would violate Euler's polyhedron formula ($V-E+F=2$). However, this can be done at the cost of introducing defects. We implement this approach using a series of polyhedra where the faces are quadrilaterals and triangles. As we move down the series, the quadrilateral faces grow in number while the number of triangles remains fixed at eight. These triangles can be viewed as defect rings that arise due to curvature of the underlying space. Crucially, every cluster in this series maintains a coordination-number of four. This allows us to define ice models and to study their physics.

Fig.~\ref{fig.clusters} shows the first four polyhedra in the series, in order from left to right. At the far left, we have an octahedron. Each subsequent polyhedron is obtained by `rectification'\cite{Conway2008} of its immediate predecessor. Rectification is a geometric operation where new vertices are defined at the bond centres of the previous polyhedron. These vertices are then connected by faces so as to cut off the old vertices. This can be viewed as cutting off the extremities of the old cluster by sawing through planes that connect bond-midpoints. In Conway polyhedron notation, rectification is denoted as an `ambo' operation\cite{Conway2008}. Each cluster can be thought of as a planar graph or Schlegel diagram. In this point of view, the rectification operation is equivalent to constructing a medial graph.

Every cluster in Fig.~\ref{fig.clusters} has eight triangular faces. It can be easily seen that subsequenct clusters in this series will also have precisely eight triangular faces. 
The number of quadrilateral faces grows as we move down the series. The octahedron has no quadrilaterals; the cuboctahedron has 6; the rhombicuboctahedron 18 and the expanded cuboctahedron 42. If the series were to continue, the limiting cluster would provide an infinite square grid with the topology of a sphere, with eight `defects' in the form of triangular faces.

\subsection{Enumerating ice configurations}
\label{sec.enumerate}
We define ice models by drawing arrows on bonds and enforcing the two-in-two-out rule at each vertex. We take a numerical approach, representing each ice configuration as a directed graph. On the octahedron and cuboctahedron, all possible configurations can be enumerated by brute force. On larger clusters, we use specialized algorithms where we enumerate configurations on smaller patches and `stitch' them together -- see Appendix.~\ref{app.stitch}.

\begin{table*}
\begin{tabular}{|c|c|c|c|c|c|c|}
\hline
\textbf{Cluster} & \textbf{Vertices} & \textbf{Ice configurations} & $\mathbf{N_{tri.}}$ & $\mathbf{N_{quad.}}$ & $\mathbf{f_{tri.}}$ & $\mathbf{f_{quad.}}$\\
\hline\hline
Octahedron & 6 & 38 & 20 & - & 52.63\% & -\\
\hline
Cuboctahedron & 12 & 600 & 384 & 192 & 64\% & 32\% \\ 
\hline
Rhombicuboctahedron & 24 & 118,976 &  69776 & 
$\begin{array}{c}
43192\\
56416
\end{array}$
& 58.64\% & 
$\begin{array}{c}
36.30\%, \\47.42\% \end{array}$\\ 
\hline
Expanded cuboctahedron & 48 & 4,292,378,946 & 2,505,964,740 & $\begin{array}{c}
1,664,338,936\\ 
1,909,987,152\\ 
1,802,996,960
\end{array}$ & 58.38\% & 
$\begin{array}{c}
38.77\%\\
44.50\%\\ 
42.00\% 
\end{array}$
 \\
\hline\hline
Pauling-like estimate & - & - & - & - & 57.14\% & 39.02\% \\
\hline
\end{tabular}
\caption{Numerical results from enumerating ice configurations. The `ice configurations' column gives the number of ice-rule-satisfying configurations on each cluster. $N_{tri.}$ ($N_{quad.}$) represents the number of configurations in which a given triangle (quadrilateral) is flippable. In clusters with symmetry-distinct quadrilaterals, $N_{quad.}$ takes multiple values. They are listed in order of distance from triangles (those that share an edge, followed by those that share a vertex and finally those that are separated). The column $f_{tri.}$ lists the flippability of the triangle ring, while $f_{quad.}$ denotes the flippability of quadrilaterals. Flippability values are truncated to two decimal places. The last row is the result from the Pauling-like argument of Eq.~\ref{eq.fM}.}
\label{tab.clusters}
\end{table*}

The total number of ice configurations on each cluster is shown in Tab.~\ref{tab.clusters}. The number grows exponentially as we move down the series, starting from 38 for the octahedron and exceeding 4 billion in the expanded cuboctahedron. The exponential growth is consistent with Pauling's entropy expression which predicts $N \sim (3/2)^V$ where $V$ is the number of vertices. However, the precise numerical values are not in good agreement with the Pauling estimate. This could be due to variations in local environments (e.g., triangles vs. quadrilaterals).

\subsection{Flippability}
As we enumerate ice configurations on each cluster, we examine whether a given ring is flippable, i.e., whether its internal arrows are pointed in the same sense. 
We define flippability of the ring as the fraction of configurations in which this ring is flippable. For example, the cuboctahedron has 600 ice configurations. Focussing on one triangular ring, we find that it is flippable in 384 of the 600 configurations. We define $f_{tri.}$, the flippability of the triangle, as 384/600 or 64\%.

Numerical results for flippability are shown in Tab.~\ref{tab.clusters}. On each cluster, we show flippability for every symmetry-distinct ring. For example, as the eight triangles in each cluster are symmetry-related, they must all have the same flippability. It suffices to calculate one triangle-flippability, listed as $f_{tri.}$. In the rhombicuboctahedron, we have two symmetry-distinct quadrilateral rings. Twelve quadrilateral faces share an edge with a triangle, while six share a corner. They have distinct flippabilities, both listed under the $f_{quad.}$ column. The expanded cuboctahedron has three distinct quadrilateral rings. Twenty four quadrilaterals share an edge with a triangle. Twelve share a corner. Finally, six are entirely separated from triangles. This results in three values of quadrilateral flippability, as shown in the $f_{quad.}$ column.

Qualitatively, the numerical results show that triangles are more flippable than quadrilaterals. Quantitatively, the numerically obtained flippabilities show broad agreement with Eq.~\ref{eq.fM}, the result of the Pauling-like single-ring argument. For example, Eq.~\ref{eq.fM} yields $\sim$57.14\% for the flippability of a triangle. This can be compared to $\sim$58.38\% in the expanded cuboctahedron. 
For a quadrilateral ring, the Pauling-like argument predicts $\sim$39.02\%. This is comparable to (38.77\%, 44.50\%, 42.00\%) on the three symmetry-distinct quadrilaterals of the expanded cuboctahedron. Crucially, the quadrilaterals are always less flippable than the triangles. Eq.~\ref{eq.fM} predicts $f_{quad.}/f_{tri.} \sim 68\%$. This compares favourably with the numerical values in Tab.~\ref{tab.clusters}.

\section{Quantum tunnelling in spherical square ice }
\label{sec.quantum}
As a minimal model for dynamics on the clusters of Fig.~\ref{fig.clusters}, we study a model with `tunnelling' processes that flip rings. We invoke a Hamiltonian with local flips on the smallest rings (triangles and quadrilaterals). The Hamiltonian takes the form
\begin{align}
\nonumber \hat{H} = &-t_\triangle \sum_{\triangle} 
\Big\{ \big\vert  \begin{gathered}\includegraphics[width=0.2in]{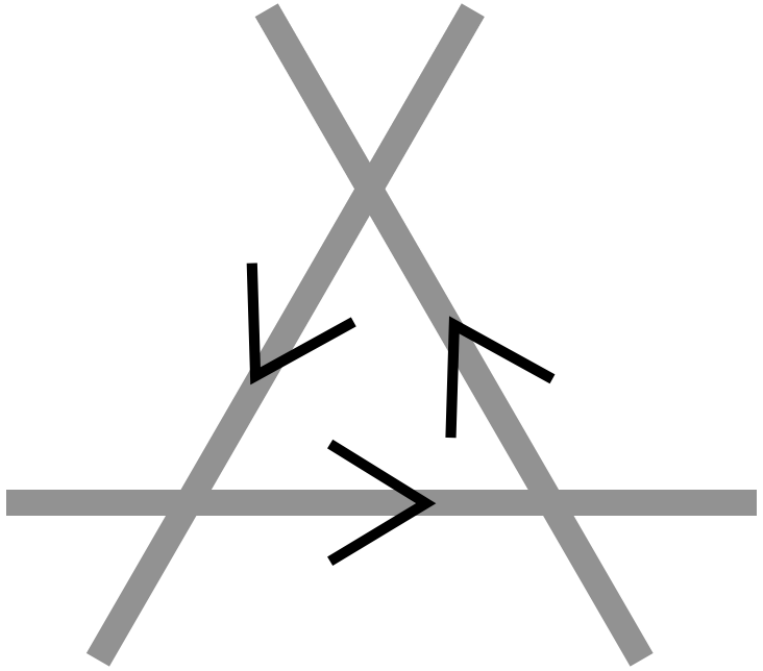} \end{gathered}
\big\rangle \big\langle \begin{gathered} \includegraphics[width=0.2in]{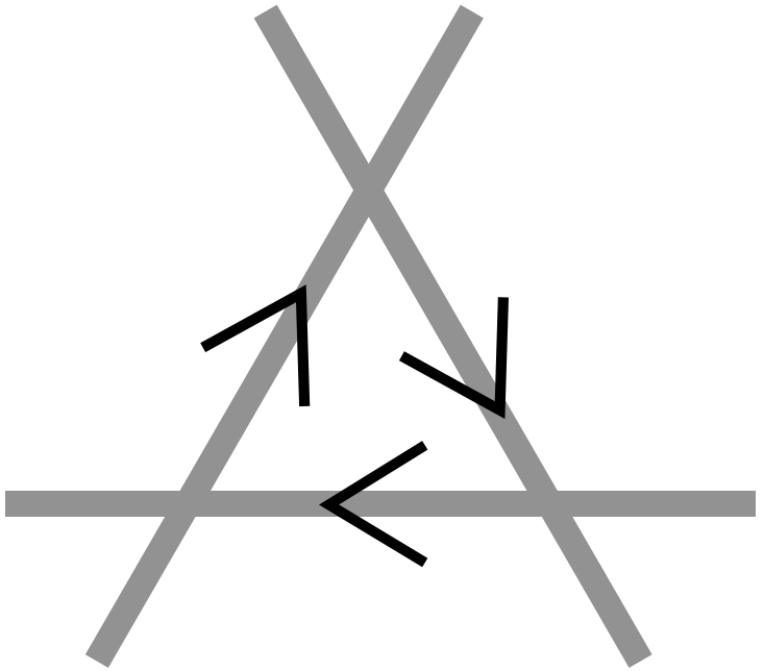}\end{gathered} \big\vert 
+\big\vert  \begin{gathered}\includegraphics[width=0.2in]{tri2} \end{gathered}
\big\rangle \big\langle \begin{gathered} \includegraphics[width=0.2in]{tri1}\end{gathered} \big\vert 
\Big\}\\
&-t_\square \sum_{\square} 
\Big\{ \big\vert  \begin{gathered}\includegraphics[width=0.2in]{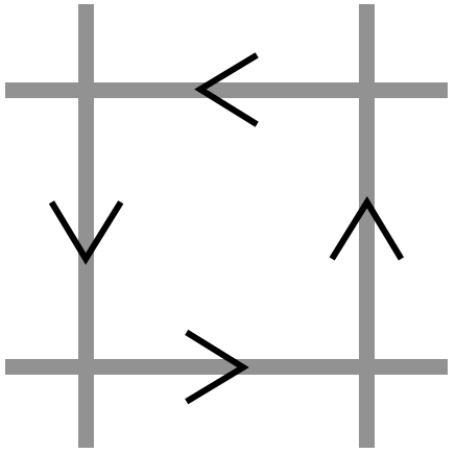} \end{gathered}
\big\rangle \big\langle \begin{gathered} \includegraphics[width=0.2in]{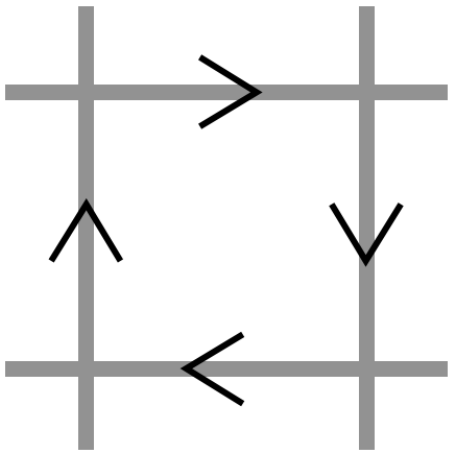}\end{gathered} \big\vert 
+\big\vert  \begin{gathered}\includegraphics[width=0.2in]{sq2} \end{gathered}
\big\rangle \big\langle \begin{gathered} \includegraphics[width=0.2in]{sq1}\end{gathered} \big\vert 
 \Big\},
\end{align}
where the summations run over all rings in the cluster. Each term in the Hamiltonian flips arrows within a single ring \textit{if} the ring is initially in a flippable state. All external arrows are left untouched. As all triangles are symmetry equivalent, they must have the same flipping amplitude, denoted as $t_\triangle$. For simplicity, we ignore symmetry distinctions between quadrilateral rings and assign the same flipping amplitude to each one, $t_\square$.

We may expect $t_\triangle$ to be larger in magnitude than $t_\square$. This expectation stems from microscopic studies where flipping processes arise as terms in a perturbative expansion\cite{Hermele2004,CastroNeto2006,Henry2014}. Flipping a ring with $M$ edges requires $M$ arrow-flips, involving $M-1$ virtual states. Larger rings have lower flipping amplitudes as they involve a larger number of ice-rule-violating virtual states. This leads to a \textit{kinetic} effect whereby flipping amplitudes decay with ring size. In particular, this will lead to triangles showing a greater degree of dynamics than quadrilaterals. Here, we are interested in a distinct \textit{statistical} effect where differences in flippability lead to a hierarchy in dynamics. In order to extract the statistical effect alone, we study a simplified model with $t_\triangle =t_\square$ (set to unity). We will see below that triangles are more dynamic even though they are assigned the same flipping amplitude as quadrilaterals.

For each cluster, we express this Hamiltonian as an $N \times N$ matrix, where $N$ is the number of ice configurations (see Tab.~\ref{tab.clusters}). We diagonalize this matrix. For the octahedron and cuboctahedron, we perform full diagonalization, finding all eigenstates. As the Hilbert space for the rhombicuboctahedron is too large for this approach, we take advantage of the sparse character of the matrix which allows Lanczos diagonalization to find the lowest few eigenstates. For the expanded cuboctahedron and higher members of the series, the Hilbert space dimension is too large to be accessible within our computational resources. 

From the octahedron to the rhombicuboctahedron, the Hilbert space is fully connected and bipartite. This leads to a spectral reflection symmetry: an eigenstate with energy $E$ is associated with a partner with energy $-E$, see App.~\ref{app.bipartite}. Below, we focus on the ground state and its properties. 

\subsection{Nature of ground state}

\begin{figure}
\includegraphics[width=\columnwidth]{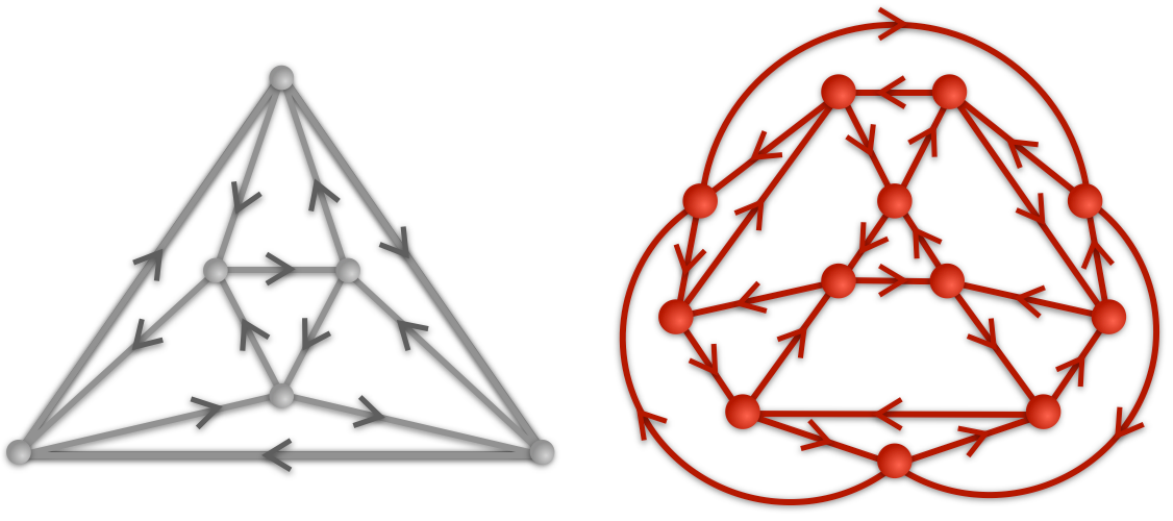}
\caption{A maximally flippable configuration on the octahedron (left) and cuboctahedron (right), both shown as Schlegel diagrams. Each face has a well-defined circulation, with adjacent faces having opposite circulation. Each cluster has one other maximally flippable configuration, obtained by reversing all the arrows in the figure.   }
\label{fig.maxflip}
\end{figure}

As the Hilbert space is fully connected, the ground state of each cluster is a linear superposition of \textit{all} ice-rule-satisfying configurations. However, the amplitudes are not uniform. In fact, the largest amplitudes occur at two configurations that are `maximally flippable'. These two configurations can be constructed as follows. We first note that the clusters in Fig.~\ref{fig.clusters} (as well as all subsequent clusters in the series) are face-bipartite. That is, their faces (triangle/quadrilateral rings) can be grouped into two families such that each edge connects a face from one family to a face from the other. To have maximal flippability, we designate one family of faces as `outward' and the other as `inward'. On the edges surrounding each outward face, we draw arrows so that the circulation points outwards from the centre of the cluster. This automatically ensures that the inward faces have the opposite circulation. Fig.~\ref{fig.maxflip} shows the resulting configurations in the octahedron and cuboctahedron. There are two such maximally flippable configurations on each cluster, as we may choose one of two families to be outward. These configurations are highly conducive to tunnelling as they allow for the most number of flips.

On the octahedron, the two maximally flippable configurations contribute $\sim$18.18\% of the weight in the ground state. On the cuboctahedron and rhombicuboctahedron, the contributions are $\sim$1.66\% and $\sim$0.06\%. These represent large contributions. For example, in the rhombicuboctahedron, the fraction of maximally flippable states is 2/118,976$\sim$0.0017\%. Yet, these states contribute $\sim$0.06\% of the ground state weight. Due to this high overlap, correlations in the ground states reflect those of the maximally flippable states. For example, if one face is flippable, there is a high likelihood for the neighbouring faces to be flippable but with opposite circulation.

\subsection{Ring contributions to ground state energy}

As the Hamiltonian is a sum of ring-flipping terms, we may examine the contribution of each term separately to the ground state energy. The ground state energy for each cluster is shown in Tab.~\ref{tab.quantum}. On a given triangle, we define $\la H_{tri.} \ra_0 =  \Big\la \Big\{ \big\vert  \begin{gathered}\includegraphics[width=0.2in]{tri1} \end{gathered}
\big\rangle \big\langle \begin{gathered} \includegraphics[width=0.2in]{tri2}\end{gathered} \big\vert   + h.c.\Big\} \Big\ra$, the expectation value of the local flipping operator in the ground state of the cluster. This term represents local resonance within a triangle, encoding the triangle's contribution to the ground state energy. Similarly, on a given quadrilateral, we define $\la H_{quad.} \ra_0  =  \la \Big\{ \big\vert  \begin{gathered}\includegraphics[width=0.2in]{sq1} \end{gathered}\big\rangle \big\langle \begin{gathered} \includegraphics[width=0.2in]{sq2}\end{gathered} \big\vert   + h.c.\Big\} \ra$. In clusters with multiple symmetry-distinct quadrilaterals, we obtain multiple values of $\la H_{quad.} \ra_0$. In other words, each symmetry-distinct quadrilateral provides a different contribution to the ground state energy.

As shown in Tab.~\ref{tab.quantum}, in all clusters, the largest contribution to the ground state energy is from triangles. In the rhombicuboctahedron, we have 8 triangles and 18 quadrilaterals. Despite being fewer in number, the triangles contribute $\sim 53 \%$ of the ground state energy. 
 
\begin{table}
\begin{tabular}{|c|c|c|c|}
\hline
{Cluster} & $E_{g.s}$ & $\la H_{tri.}\ra_0 $ & $\la H_{quad.}\ra_0$\\
\hline\hline
Octahedron & -4.6904 & -0.5863 & - \\
Cuboctahedron &  -8.2269 & -0.9532 & -0.1002 \\ 
Rhombicuboctahedron & -14.4446 & -0.9575 & -0.0933, -0.9441 \\
\hline
\end{tabular}
\caption{
Ground state contributions on clusters. The column $E_{g.s.}$ represents the ground state energy of the Hamiltonian defined in the main text. The column, $\la H_{tri.}\ra_0 $, represents the resonance contribution of a single triangular plaquette to the ground state energy. Finally,  $\la H_{quad.}\ra_0 $, represents the resonance contribution of a single quadrilateral. In clusters with symmetry-distinct quadrilaterals, multiple values are shown for $\la H_{quad.}\ra_0 $.
}
\label{tab.quantum}
\end{table}

\section{Discussion}

Our key result is a \textit{statistical} effect whereby the flippability of a ring decreases with polygon size. This is distinct from the well-known \textit{kinetic} effect whereby flipping amplitudes decay for large ring sizes. The kinetic effect can be understood from a perturbative expansion where flipping a larger ring is a higher order process\cite{Hermele2004,CastroNeto2006,Henry2014}. In contrast, the statistical effect originates from the geometry of rings and the ice rules themselves. In any ice system, the kinetic and statistical effects will act in conjunction to enhance (suppress) local dynamics at smaller (larger) rings.

In Sec.~\ref{sec.enumerate} above, we have enumerated ice configurations on spherical clusters. Our results could motivate experiments on several fronts. Recent experiments with nanomagnets have created a three dimensional array in the shape of a buckyball\cite{Donnelly2015}. It may be possible to synthesize polyhedra such as the ones we have considered in Fig.~\ref{fig.clusters}. Experiments could probe the statistics of local orientations around defects\cite{Perrin2016} as well as the degree of local dynamics. 
Quantum annealing circuits have simulated spin ice\cite{King2021} by designing two-qubit couplings to mimic a square grid. A suitable choice of couplings may be able to encode polyhedral ice. Defect rings may play a strong role in annealing, e.g., with triangles flipping more easily than quadrilaterals. Water molecules enclosed between graphene sheets have been shown to realize square ice\cite{Algara2015}. Hydrophobic structures with spherical geometry could realize polyhedral ice.

Previous theoretical studies of quantum dynamics have used Hamiltonians with ring-flip processes\cite{Benton2016,Kondakor2023}. In Sec.~\ref{sec.quantum}, we have taken the same approach, using exact diagonalization to solve for the ground state on each cluster. For clusters with larger sizes and different curved topologies, quantum Monte Carlo methods may help in describing low energy properties. A suitable field theory approach may be able to account for defects induced by curvature. Such an approach can describe large systems where numerical enumeration is not possible. Defect rings may have interesting consequences. For example, curvature-free models host phase transitions upon varying relative energies of the six local configurations at each site\cite{Kondakor2023}. Defect rings may smoothen such transitions or alter their critical exponents.

\acknowledgments
We thank Karlo Penc and M\'ark Kond\'akor for many helpful discussions and critical feedback. This work was supported by the Natural Sciences and Engineering Research Council of Canada through Discovery Grant 2022-05240.\\

\appendix
\section{Ice configurations on a polygon with $M$ vertices}
\label{app.polyM}
Consider a polygon with M sides. At each vertex, we have two internal arrows that connect to neighbouring vertices. We also have two external arrows that point outwards. If the internal arrows are in a one-in, one-out configuration, we have two possibilities for the external arrows -- we may choose one of them to be outwards and the other to be inwards. If the internal arrows are in a two-out setting, the external arrows are forced to be two-in. Likewise, if the internal arrows are two-in, the external arrows are forced to be two-out. 

To enumerate the number of ice configurations, we consider a toy problem where we build a linear arrangement, adding one bond at a time. A similar setup was considered by Nagle as a stepping stone to residual entropy on lattices\cite{Nagle1966}. On the first bond, we may draw an arrow in one of two possible ways. We then add a second bond -- with two external arrows at the junction between the first and the second bonds. The arrow on the second bond may point in either direction. Depending on the arrows on the first and second bonds, the external arrows can have one of two possibilities. We proceed to add bonds in this fashion. Below, we formulate a recursion relation for the number of configurations. 

Suppose we have enumerated all possible configurations for an arrangement with $M$ bonds. We classify these configurations into two families. In one family, the first and the last internal bonds are parallel. We denote the number of configurations in this family as $g_M$. In the other, the first and the last are opposite. The number of configurations in this family is denoted as $f_M$. In the simplest case, we may take $M=2$. By explicit enumeration, we have ($g_2=4$, $f_2=2$), ($g_3 = 10$, $f_3=8$), etc.  

We now add a new vertex at the end of the $M^\mathrm{th}$ bond. We then add the $(M+1)^\mathrm{th}$ bond as well as two external arrows  at the new vertex. We have two choices for arrow direction on the $(M+1)^\mathrm{th}$ bond: parallel or opposite to that on the immediately preceding ($M^\mathrm{th}$) bond. If it is parallel, we have two possibilities for the external arrows on the new vertex.
 If it is opposite, we have only one. These possibilities arise from the ice rule constraint on the newly added vertex. Using these arguments, we write
 \bea
 g_{M+1} = 2 g_M + f_M;~~f_{M+1} = 2f_M + g_M.
 \label{eq.recursion}
 \eea
From the values of $g_2$, $f_2$, $g_3$, and $f_3$ above, we see that $g_{M} -f_M =2$ for $M=2,3$. Crucially, if $g_M - f_M =2$, the recursion relations of Eq.~\ref{eq.recursion} guarantee $g_{M+1} - f_{M+1} =2$. We conclude that $g_M = f_M+2$ for all $M\geq 2$. The expression for $g_{M+1}$ given above can now be rewritten as $g_{M+1} = 3g_M-2$. We have arrived at a recursion relation for $g_M$, with no reference to $f_M$. It can be easily seen that this relation is satisfied by $g_M = 1+3^{M-1}$. 

Finally, we argue that $g_{M+1}$ is the number of ice-rule satisfying configurations on a polygon with $M$ sides. Thus far, we have enumerated arrow configurations on a linear chain of bonds. By definition, $g_{M+1}$ is the number of configurations of $M+1$ bonds such that the first and the last have parallel (internal) arrows. This is equivalent to stipulating periodic boundary conditions in an $M$-bond problem; in other words, it constructs a polygon with $M$ sides. The number of ice-rule satisfying configurations on the polygon with $M$ sides is $g_{M+1} = 1+3^M$.

\section{Patch stitching algorithm}
\label{app.stitch}
\begin{figure}
\includegraphics[width=1.5in]{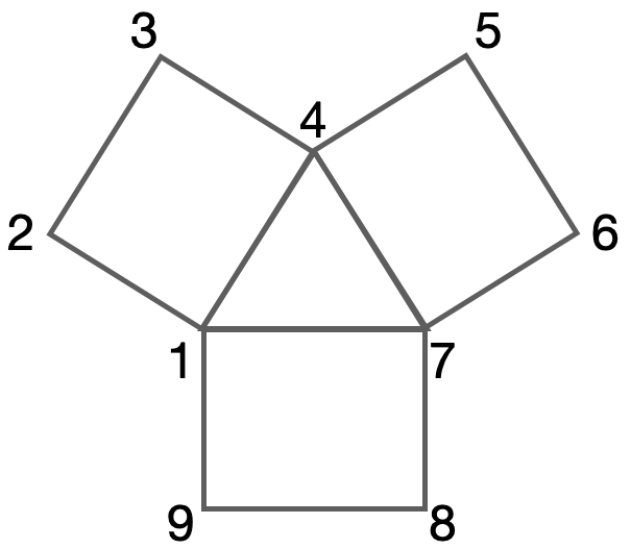}
\includegraphics[width=1.5in]{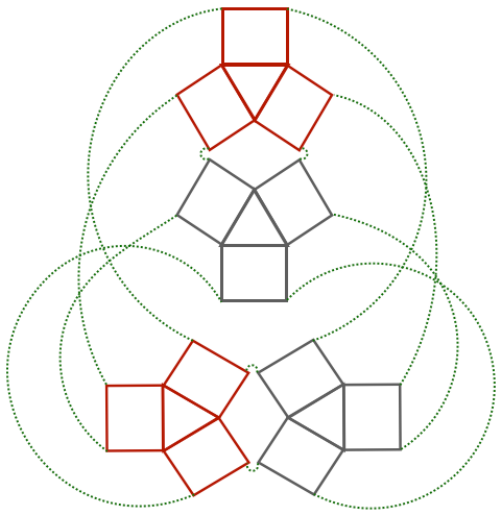}
\caption{Left: A patch that serves as a building block for the rhombicuboctahedron and the expanded cuboctahedron. Right: The construction of a rhombicuboctahedron by `stitching' four patches together. Each dotted line represents a stitching operation where a pair of vertices from different patches are identified. }
\label{fig.patch}
\end{figure}

A brute force approach does not work when enumerating ice configurations on larger clusters. With 48 edges, the rhombicuboctahedron has for $2^{48}$ possibilities that must be checked against ice rules. The expanded cuboctahedron has $2^{96}$. Instead, we use a divide and conquer approach. We identify the 9-vertex patch shown in Fig.~\ref{fig.patch} as a building block of these two clusters. Due its small size, its ice-configurations can be enumerated easily -- with 224 possibilities. There are six corner vertices where two edges meet -- the ice-rules are not enforced at these vertices.    

The rhombicuboctahedron can now be constructed by `stitching' four such patches, as shown in Fig.~\ref{fig.patch}. Each stitching operation involves two corner vertices located on two distinct patches. 
They are fused into a single vertex \textit{if} allowed by the ice rule. That is, if two edges from one patch and the two from the other combine to give a two-in-two-out configuration, the vertices can be fused. If the rule is not satisfied (e.g., if we have two incoming arrows on both patches, leading to four incoming arrows in total), the stitching operation terminates, discarding the current configuration. The rhombicuboctahedron is obtained by stitching four patches, while the expanded cuboctahedron requires eight patches. In each case, it is efficient to proceed by stitching two similar units at a time. For example, two patches can be stitched to form a bi-patch. Two bi-patches can be then stitched to form a rhombicuboctahedron.

\section{Bipartite character of tunnelling Hamiltonian}
\label{app.bipartite}

On the four clusters shown in Fig.~\ref{fig.clusters}, ice configurations separate into two families. Flipping processes always connect an element from one family to one in the other. This bipartite character leads to a spectral reflection symmetry\cite{Iadecola2018}: an eigenstate with energy $E$ is associated with a partner with energy $-E$. This bipartite character can be understood by comparison with a reference configuration. For example, on the octahedron, we compare any given ice configuration with a fixed reference configuration (say a maximally flippable configuration). We count the number of arrows that are parallel between the given configuration and the reference. Based on this number, we classify the configuration as even or odd. As a flip will alter three arrows at a time, it will necessarily connect even to odd and vice versa. On the cuboctahedron, rhombicuboctahedron and expanded cuboctahedron, we cannot take the same approach as we have quadrilateral faces. A quadrilateral flip will change four arrows and may connect even-to-even or odd-to-odd. However, on each cluster, it is possible to construct a subgraph that contains an odd number of edges within every ring. By comparison with a reference configuration on this subgraph, we may designate any given configuration as even or odd. Flipping processes will now connect even-to-odd or odd-to-even.

\bibliography{octahedron_ice}

\begin{thebibliography}{32}%
\makeatletter
\providecommand \@ifxundefined [1]{%
 \@ifx{#1\undefined}
}%
\providecommand \@ifnum [1]{%
 \ifnum #1\expandafter \@firstoftwo
 \else \expandafter \@secondoftwo
 \fi
}%
\providecommand \@ifx [1]{%
 \ifx #1\expandafter \@firstoftwo
 \else \expandafter \@secondoftwo
 \fi
}%
\providecommand \natexlab [1]{#1}%
\providecommand \enquote  [1]{``#1''}%
\providecommand \bibnamefont  [1]{#1}%
\providecommand \bibfnamefont [1]{#1}%
\providecommand \citenamefont [1]{#1}%
\providecommand \href@noop [0]{\@secondoftwo}%
\providecommand \href [0]{\begingroup \@sanitize@url \@href}%
\providecommand \@href[1]{\@@startlink{#1}\@@href}%
\providecommand \@@href[1]{\endgroup#1\@@endlink}%
\providecommand \@sanitize@url [0]{\catcode `\\12\catcode `\$12\catcode
  `\&12\catcode `\#12\catcode `\^12\catcode `\_12\catcode `\%12\relax}%
\providecommand \@@startlink[1]{}%
\providecommand \@@endlink[0]{}%
\providecommand \url  [0]{\begingroup\@sanitize@url \@url }%
\providecommand \@url [1]{\endgroup\@href {#1}{\urlprefix }}%
\providecommand \urlprefix  [0]{URL }%
\providecommand \Eprint [0]{\href }%
\providecommand \doibase [0]{http://dx.doi.org/}%
\providecommand \selectlanguage [0]{\@gobble}%
\providecommand \bibinfo  [0]{\@secondoftwo}%
\providecommand \bibfield  [0]{\@secondoftwo}%
\providecommand \translation [1]{[#1]}%
\providecommand \BibitemOpen [0]{}%
\providecommand \bibitemStop [0]{}%
\providecommand \bibitemNoStop [0]{.\EOS\space}%
\providecommand \EOS [0]{\spacefactor3000\relax}%
\providecommand \BibitemShut  [1]{\csname bibitem#1\endcsname}%
\let\auto@bib@innerbib\@empty
\bibitem [{\citenamefont {Bernal}\ and\ \citenamefont
  {Fowler}(1933)}]{BernalFowler1933}%
  \BibitemOpen
  \bibfield  {author} {\bibinfo {author} {\bibfnamefont {J.~D.}\ \bibnamefont
  {Bernal}}\ and\ \bibinfo {author} {\bibfnamefont {R.~H.}\ \bibnamefont
  {Fowler}},\ }\href {\doibase 10.1063/1.1749327} {\bibfield  {journal}
  {\bibinfo  {journal} {The Journal of Chemical Physics}\ }\textbf {\bibinfo
  {volume} {1}},\ \bibinfo {pages} {515} (\bibinfo {year} {1933})}\BibitemShut
  {NoStop}%
\bibitem [{\citenamefont {Bramwell}\ and\ \citenamefont
  {Gingras}(2001)}]{BramwellGingras2001}%
  \BibitemOpen
  \bibfield  {author} {\bibinfo {author} {\bibfnamefont {S.~T.}\ \bibnamefont
  {Bramwell}}\ and\ \bibinfo {author} {\bibfnamefont {M.~J.~P.}\ \bibnamefont
  {Gingras}},\ }\href {\doibase 10.1126/science.1064761} {\bibfield  {journal}
  {\bibinfo  {journal} {Science}\ }\textbf {\bibinfo {volume} {294}},\ \bibinfo
  {pages} {1495} (\bibinfo {year} {2001})}\BibitemShut {NoStop}%
\bibitem [{\citenamefont {Castelnovo}\ \emph {et~al.}(2012)\citenamefont
  {Castelnovo}, \citenamefont {Moessner},\ and\ \citenamefont
  {Sondhi}}]{Castelnovo2012}%
  \BibitemOpen
  \bibfield  {author} {\bibinfo {author} {\bibfnamefont {C.}~\bibnamefont
  {Castelnovo}}, \bibinfo {author} {\bibfnamefont {R.}~\bibnamefont
  {Moessner}}, \ and\ \bibinfo {author} {\bibfnamefont {S.}~\bibnamefont
  {Sondhi}},\ }\href {\doibase
  https://doi.org/10.1146/annurev-conmatphys-020911-125058} {\bibfield
  {journal} {\bibinfo  {journal} {Annual Review of Condensed Matter Physics}\
  }\textbf {\bibinfo {volume} {3}},\ \bibinfo {pages} {35} (\bibinfo {year}
  {2012})}\BibitemShut {NoStop}%
\bibitem [{\citenamefont {Bramwell}\ and\ \citenamefont
  {Harris}(2020)}]{Bramwell_2020}%
  \BibitemOpen
  \bibfield  {author} {\bibinfo {author} {\bibfnamefont {S.~T.}\ \bibnamefont
  {Bramwell}}\ and\ \bibinfo {author} {\bibfnamefont {M.~J.}\ \bibnamefont
  {Harris}},\ }\href {\doibase 10.1088/1361-648X/ab8423} {\bibfield  {journal}
  {\bibinfo  {journal} {Journal of Physics: Condensed Matter}\ }\textbf
  {\bibinfo {volume} {32}},\ \bibinfo {pages} {374010} (\bibinfo {year}
  {2020})}\BibitemShut {NoStop}%
\bibitem [{\citenamefont {Udagawa}\ and\ \citenamefont
  {Jaubert}(2021)}]{Udagawa2021}%
  \BibitemOpen
  \bibfield  {author} {\bibinfo {author} {\bibfnamefont {M.}~\bibnamefont
  {Udagawa}}\ and\ \bibinfo {author} {\bibfnamefont {L.}~\bibnamefont
  {Jaubert}},\ }\href {https://books.google.ca/books?id=JMo3zgEACAAJ} {\emph
  {\bibinfo {title} {Spin Ice}}},\ Springer Series in Solid-State Sciences\
  (\bibinfo  {publisher} {Springer International Publishing},\ \bibinfo {year}
  {2021})\BibitemShut {NoStop}%
\bibitem [{\citenamefont {Ortiz-Ambriz}\ \emph {et~al.}(2019)\citenamefont
  {Ortiz-Ambriz}, \citenamefont {Nisoli}, \citenamefont {Reichhardt},
  \citenamefont {Reichhardt},\ and\ \citenamefont {Tierno}}]{TiernoRMP2019}%
  \BibitemOpen
  \bibfield  {author} {\bibinfo {author} {\bibfnamefont {A.}~\bibnamefont
  {Ortiz-Ambriz}}, \bibinfo {author} {\bibfnamefont {C.}~\bibnamefont
  {Nisoli}}, \bibinfo {author} {\bibfnamefont {C.}~\bibnamefont {Reichhardt}},
  \bibinfo {author} {\bibfnamefont {C.~J.~O.}\ \bibnamefont {Reichhardt}}, \
  and\ \bibinfo {author} {\bibfnamefont {P.}~\bibnamefont {Tierno}},\ }\href
  {\doibase 10.1103/RevModPhys.91.041003} {\bibfield  {journal} {\bibinfo
  {journal} {Rev. Mod. Phys.}\ }\textbf {\bibinfo {volume} {91}},\ \bibinfo
  {pages} {041003} (\bibinfo {year} {2019})}\BibitemShut {NoStop}%
\bibitem [{\citenamefont {Nisoli}\ \emph {et~al.}(2013)\citenamefont {Nisoli},
  \citenamefont {Moessner},\ and\ \citenamefont {Schiffer}}]{NisoliRMP2013}%
  \BibitemOpen
  \bibfield  {author} {\bibinfo {author} {\bibfnamefont {C.}~\bibnamefont
  {Nisoli}}, \bibinfo {author} {\bibfnamefont {R.}~\bibnamefont {Moessner}}, \
  and\ \bibinfo {author} {\bibfnamefont {P.}~\bibnamefont {Schiffer}},\ }\href
  {\doibase 10.1103/RevModPhys.85.1473} {\bibfield  {journal} {\bibinfo
  {journal} {Rev. Mod. Phys.}\ }\textbf {\bibinfo {volume} {85}},\ \bibinfo
  {pages} {1473} (\bibinfo {year} {2013})}\BibitemShut {NoStop}%
\bibitem [{\citenamefont {Glaetzle}\ \emph {et~al.}(2014)\citenamefont
  {Glaetzle}, \citenamefont {Dalmonte}, \citenamefont {Nath}, \citenamefont
  {Rousochatzakis}, \citenamefont {Moessner},\ and\ \citenamefont
  {Zoller}}]{Glaetzle2014}%
  \BibitemOpen
  \bibfield  {author} {\bibinfo {author} {\bibfnamefont {A.~W.}\ \bibnamefont
  {Glaetzle}}, \bibinfo {author} {\bibfnamefont {M.}~\bibnamefont {Dalmonte}},
  \bibinfo {author} {\bibfnamefont {R.}~\bibnamefont {Nath}}, \bibinfo {author}
  {\bibfnamefont {I.}~\bibnamefont {Rousochatzakis}}, \bibinfo {author}
  {\bibfnamefont {R.}~\bibnamefont {Moessner}}, \ and\ \bibinfo {author}
  {\bibfnamefont {P.}~\bibnamefont {Zoller}},\ }\href {\doibase
  10.1103/PhysRevX.4.041037} {\bibfield  {journal} {\bibinfo  {journal} {Phys.
  Rev. X}\ }\textbf {\bibinfo {volume} {4}},\ \bibinfo {pages} {041037}
  (\bibinfo {year} {2014})}\BibitemShut {NoStop}%
\bibitem [{\citenamefont {Henley}(2010)}]{Henley2010}%
  \BibitemOpen
  \bibfield  {author} {\bibinfo {author} {\bibfnamefont {C.~L.}\ \bibnamefont
  {Henley}},\ }\href {\doibase 10.1146/annurev-conmatphys-070909-104138}
  {\bibfield  {journal} {\bibinfo  {journal} {Annual Review of Condensed Matter
  Physics}\ }\textbf {\bibinfo {volume} {1}},\ \bibinfo {pages} {179} (\bibinfo
  {year} {2010})}\BibitemShut {NoStop}%
\bibitem [{\citenamefont {Caravelli}\ \emph {et~al.}(2021)\citenamefont
  {Caravelli}, \citenamefont {Saccone},\ and\ \citenamefont
  {Nisoli}}]{Caravelli2021}%
  \BibitemOpen
  \bibfield  {author} {\bibinfo {author} {\bibfnamefont {F.}~\bibnamefont
  {Caravelli}}, \bibinfo {author} {\bibfnamefont {M.}~\bibnamefont {Saccone}},
  \ and\ \bibinfo {author} {\bibfnamefont {C.}~\bibnamefont {Nisoli}},\ }\href
  {\doibase 10.1098/rspa.2021.0108} {\bibfield  {journal} {\bibinfo  {journal}
  {Proceedings of the Royal Society A: Mathematical, Physical and Engineering
  Sciences}\ }\textbf {\bibinfo {volume} {477}},\ \bibinfo {pages} {20210108}
  (\bibinfo {year} {2021})}\BibitemShut {NoStop}%
\bibitem [{\citenamefont {Pauling}(1935)}]{Pauling1935}%
  \BibitemOpen
  \bibfield  {author} {\bibinfo {author} {\bibfnamefont {L.}~\bibnamefont
  {Pauling}},\ }\href {\doibase 10.1021/ja01315a102} {\bibfield  {journal}
  {\bibinfo  {journal} {Journal of the American Chemical Society}\ }\textbf
  {\bibinfo {volume} {57}},\ \bibinfo {pages} {2680} (\bibinfo {year}
  {1935})}\BibitemShut {NoStop}%
\bibitem [{\citenamefont {Lieb}(1967)}]{Lieb1967}%
  \BibitemOpen
  \bibfield  {author} {\bibinfo {author} {\bibfnamefont {E.~H.}\ \bibnamefont
  {Lieb}},\ }\href {\doibase 10.1103/PhysRevLett.18.692} {\bibfield  {journal}
  {\bibinfo  {journal} {Phys. Rev. Lett.}\ }\textbf {\bibinfo {volume} {18}},\
  \bibinfo {pages} {692} (\bibinfo {year} {1967})}\BibitemShut {NoStop}%
\bibitem [{\citenamefont {Chaplin}()}]{Chaplin}%
  \BibitemOpen
  \bibfield  {author} {\bibinfo {author} {\bibfnamefont {M.}~\bibnamefont
  {Chaplin}},\ }\href@noop {} {\enquote {\bibinfo {title} {Water structure and
  science},}\ }\bibinfo {howpublished}
  {\url{https://water.lsbu.ac.uk/water/water_structure_science.html}},\
  \bibinfo {note} {accessed: 2024-06-24}\BibitemShut {NoStop}%
\bibitem [{\citenamefont {Petrenko}\ and\ \citenamefont
  {Whitworth}(1999)}]{Petrenko1999}%
  \BibitemOpen
  \bibfield  {author} {\bibinfo {author} {\bibfnamefont {V.}~\bibnamefont
  {Petrenko}}\ and\ \bibinfo {author} {\bibfnamefont {R.}~\bibnamefont
  {Whitworth}},\ }\href {https://books.google.ca/books?id=oC941a8lXWIC} {\emph
  {\bibinfo {title} {Physics of Ice}}}\ (\bibinfo  {publisher} {OUP Oxford},\
  \bibinfo {year} {1999})\BibitemShut {NoStop}%
\bibitem [{\citenamefont {Bove}\ \emph {et~al.}(2009)\citenamefont {Bove},
  \citenamefont {Klotz}, \citenamefont {Paciaroni},\ and\ \citenamefont
  {Sacchetti}}]{Bove2009}%
  \BibitemOpen
  \bibfield  {author} {\bibinfo {author} {\bibfnamefont {L.~E.}\ \bibnamefont
  {Bove}}, \bibinfo {author} {\bibfnamefont {S.}~\bibnamefont {Klotz}},
  \bibinfo {author} {\bibfnamefont {A.}~\bibnamefont {Paciaroni}}, \ and\
  \bibinfo {author} {\bibfnamefont {F.}~\bibnamefont {Sacchetti}},\ }\href
  {\doibase 10.1103/PhysRevLett.103.165901} {\bibfield  {journal} {\bibinfo
  {journal} {Phys. Rev. Lett.}\ }\textbf {\bibinfo {volume} {103}},\ \bibinfo
  {pages} {165901} (\bibinfo {year} {2009})}\BibitemShut {NoStop}%
\bibitem [{\citenamefont {Drechsel-Grau}\ and\ \citenamefont
  {Marx}(2014)}]{Marx2014}%
  \BibitemOpen
  \bibfield  {author} {\bibinfo {author} {\bibfnamefont {C.}~\bibnamefont
  {Drechsel-Grau}}\ and\ \bibinfo {author} {\bibfnamefont {D.}~\bibnamefont
  {Marx}},\ }\href {\doibase 10.1103/PhysRevLett.112.148302} {\bibfield
  {journal} {\bibinfo  {journal} {Phys. Rev. Lett.}\ }\textbf {\bibinfo
  {volume} {112}},\ \bibinfo {pages} {148302} (\bibinfo {year}
  {2014})}\BibitemShut {NoStop}%
\bibitem [{\citenamefont {Yen}\ and\ \citenamefont {Gao}(2015)}]{Yen2015}%
  \BibitemOpen
  \bibfield  {author} {\bibinfo {author} {\bibfnamefont {F.}~\bibnamefont
  {Yen}}\ and\ \bibinfo {author} {\bibfnamefont {T.}~\bibnamefont {Gao}},\
  }\bibfield  {booktitle} {\emph {\bibinfo {booktitle} {The Journal of Physical
  Chemistry Letters}},\ }\href {\doibase 10.1021/acs.jpclett.5b00797}
  {\bibfield  {journal} {\bibinfo  {journal} {The Journal of Physical Chemistry
  Letters}\ }\textbf {\bibinfo {volume} {6}},\ \bibinfo {pages} {2822}
  (\bibinfo {year} {2015})}\BibitemShut {NoStop}%
\bibitem [{\citenamefont {Kamb}\ \emph {et~al.}(1967)\citenamefont {Kamb},
  \citenamefont {Prakash},\ and\ \citenamefont {Knobler}}]{Kamb1967}%
  \BibitemOpen
  \bibfield  {author} {\bibinfo {author} {\bibfnamefont {B.}~\bibnamefont
  {Kamb}}, \bibinfo {author} {\bibfnamefont {A.}~\bibnamefont {Prakash}}, \
  and\ \bibinfo {author} {\bibfnamefont {C.}~\bibnamefont {Knobler}},\ }\href
  {\doibase 10.1107/S0365110X67001409} {\bibfield  {journal} {\bibinfo
  {journal} {Acta Crystallographica}\ }\textbf {\bibinfo {volume} {22}},\
  \bibinfo {pages} {706} (\bibinfo {year} {1967})}\BibitemShut {NoStop}%
\bibitem [{\citenamefont {Pirzadeh}\ and\ \citenamefont
  {Kusalik}(2011)}]{Pirzadeh2010}%
  \BibitemOpen
  \bibfield  {author} {\bibinfo {author} {\bibfnamefont {P.}~\bibnamefont
  {Pirzadeh}}\ and\ \bibinfo {author} {\bibfnamefont {P.~G.}\ \bibnamefont
  {Kusalik}},\ }\href {\doibase 10.1021/ja109273m} {\bibfield  {journal}
  {\bibinfo  {journal} {Journal of the American Chemical Society}\ }\textbf
  {\bibinfo {volume} {133}},\ \bibinfo {pages} {704} (\bibinfo {year}
  {2011})},\ \bibinfo {note} {pMID: 21190379}\BibitemShut {NoStop}%
\bibitem [{\citenamefont {Rosu-Finsen}\ \emph {et~al.}(2023)\citenamefont
  {Rosu-Finsen}, \citenamefont {Davies}, \citenamefont {Amon}, \citenamefont
  {Wu}, \citenamefont {Sella}, \citenamefont {Michaelides},\ and\ \citenamefont
  {Salzmann}}]{Salzmann2015}%
  \BibitemOpen
  \bibfield  {author} {\bibinfo {author} {\bibfnamefont {A.}~\bibnamefont
  {Rosu-Finsen}}, \bibinfo {author} {\bibfnamefont {M.~B.}\ \bibnamefont
  {Davies}}, \bibinfo {author} {\bibfnamefont {A.}~\bibnamefont {Amon}},
  \bibinfo {author} {\bibfnamefont {H.}~\bibnamefont {Wu}}, \bibinfo {author}
  {\bibfnamefont {A.}~\bibnamefont {Sella}}, \bibinfo {author} {\bibfnamefont
  {A.}~\bibnamefont {Michaelides}}, \ and\ \bibinfo {author} {\bibfnamefont
  {C.~G.}\ \bibnamefont {Salzmann}},\ }\href {\doibase 10.1126/science.abq2105}
  {\bibfield  {journal} {\bibinfo  {journal} {Science}\ }\textbf {\bibinfo
  {volume} {379}},\ \bibinfo {pages} {474} (\bibinfo {year}
  {2023})}\BibitemShut {NoStop}%
\bibitem [{\citenamefont {Conway}\ \emph {et~al.}(2008)\citenamefont {Conway},
  \citenamefont {Burgiel},\ and\ \citenamefont {Goodman-Strauss}}]{Conway2008}%
  \BibitemOpen
  \bibfield  {author} {\bibinfo {author} {\bibfnamefont {J.}~\bibnamefont
  {Conway}}, \bibinfo {author} {\bibfnamefont {H.}~\bibnamefont {Burgiel}}, \
  and\ \bibinfo {author} {\bibfnamefont {C.}~\bibnamefont {Goodman-Strauss}},\
  }\href {https://books.google.ca/books?id=EtQCk0TNafsC} {\emph {\bibinfo
  {title} {The Symmetries of Things}}},\ AK Peters/CRC Recreational Mathematics
  Series\ (\bibinfo  {publisher} {Taylor \& Francis},\ \bibinfo {year}
  {2008})\BibitemShut {NoStop}%
\bibitem [{\citenamefont {Hermele}\ \emph {et~al.}(2004)\citenamefont
  {Hermele}, \citenamefont {Fisher},\ and\ \citenamefont
  {Balents}}]{Hermele2004}%
  \BibitemOpen
  \bibfield  {author} {\bibinfo {author} {\bibfnamefont {M.}~\bibnamefont
  {Hermele}}, \bibinfo {author} {\bibfnamefont {M.~P.~A.}\ \bibnamefont
  {Fisher}}, \ and\ \bibinfo {author} {\bibfnamefont {L.}~\bibnamefont
  {Balents}},\ }\href {\doibase 10.1103/PhysRevB.69.064404} {\bibfield
  {journal} {\bibinfo  {journal} {Phys. Rev. B}\ }\textbf {\bibinfo {volume}
  {69}},\ \bibinfo {pages} {064404} (\bibinfo {year} {2004})}\BibitemShut
  {NoStop}%
\bibitem [{\citenamefont {Castro~Neto}\ \emph {et~al.}(2006)\citenamefont
  {Castro~Neto}, \citenamefont {Pujol},\ and\ \citenamefont
  {Fradkin}}]{CastroNeto2006}%
  \BibitemOpen
  \bibfield  {author} {\bibinfo {author} {\bibfnamefont {A.~H.}\ \bibnamefont
  {Castro~Neto}}, \bibinfo {author} {\bibfnamefont {P.}~\bibnamefont {Pujol}},
  \ and\ \bibinfo {author} {\bibfnamefont {E.}~\bibnamefont {Fradkin}},\ }\href
  {\doibase 10.1103/PhysRevB.74.024302} {\bibfield  {journal} {\bibinfo
  {journal} {Phys. Rev. B}\ }\textbf {\bibinfo {volume} {74}},\ \bibinfo
  {pages} {024302} (\bibinfo {year} {2006})}\BibitemShut {NoStop}%
\bibitem [{\citenamefont {Henry}\ and\ \citenamefont
  {Roscilde}(2014)}]{Henry2014}%
  \BibitemOpen
  \bibfield  {author} {\bibinfo {author} {\bibfnamefont {L.-P.}\ \bibnamefont
  {Henry}}\ and\ \bibinfo {author} {\bibfnamefont {T.}~\bibnamefont
  {Roscilde}},\ }\href {\doibase 10.1103/PhysRevLett.113.027204} {\bibfield
  {journal} {\bibinfo  {journal} {Phys. Rev. Lett.}\ }\textbf {\bibinfo
  {volume} {113}},\ \bibinfo {pages} {027204} (\bibinfo {year}
  {2014})}\BibitemShut {NoStop}%
\bibitem [{\citenamefont {Donnelly}\ \emph {et~al.}(2015)\citenamefont
  {Donnelly}, \citenamefont {Guizar-Sicairos}, \citenamefont {Scagnoli},
  \citenamefont {Holler}, \citenamefont {Huthwelker}, \citenamefont {Menzel},
  \citenamefont {Vartiainen}, \citenamefont {M\"uller}, \citenamefont {Kirk},
  \citenamefont {Gliga}, \citenamefont {Raabe},\ and\ \citenamefont
  {Heyderman}}]{Donnelly2015}%
  \BibitemOpen
  \bibfield  {author} {\bibinfo {author} {\bibfnamefont {C.}~\bibnamefont
  {Donnelly}}, \bibinfo {author} {\bibfnamefont {M.}~\bibnamefont
  {Guizar-Sicairos}}, \bibinfo {author} {\bibfnamefont {V.}~\bibnamefont
  {Scagnoli}}, \bibinfo {author} {\bibfnamefont {M.}~\bibnamefont {Holler}},
  \bibinfo {author} {\bibfnamefont {T.}~\bibnamefont {Huthwelker}}, \bibinfo
  {author} {\bibfnamefont {A.}~\bibnamefont {Menzel}}, \bibinfo {author}
  {\bibfnamefont {I.}~\bibnamefont {Vartiainen}}, \bibinfo {author}
  {\bibfnamefont {E.}~\bibnamefont {M\"uller}}, \bibinfo {author}
  {\bibfnamefont {E.}~\bibnamefont {Kirk}}, \bibinfo {author} {\bibfnamefont
  {S.}~\bibnamefont {Gliga}}, \bibinfo {author} {\bibfnamefont
  {J.}~\bibnamefont {Raabe}}, \ and\ \bibinfo {author} {\bibfnamefont {L.~J.}\
  \bibnamefont {Heyderman}},\ }\href {\doibase 10.1103/PhysRevLett.114.115501}
  {\bibfield  {journal} {\bibinfo  {journal} {Phys. Rev. Lett.}\ }\textbf
  {\bibinfo {volume} {114}},\ \bibinfo {pages} {115501} (\bibinfo {year}
  {2015})}\BibitemShut {NoStop}%
\bibitem [{\citenamefont {Perrin}\ \emph {et~al.}(2016)\citenamefont {Perrin},
  \citenamefont {Canals},\ and\ \citenamefont {Rougemaille}}]{Perrin2016}%
  \BibitemOpen
  \bibfield  {author} {\bibinfo {author} {\bibfnamefont {Y.}~\bibnamefont
  {Perrin}}, \bibinfo {author} {\bibfnamefont {B.}~\bibnamefont {Canals}}, \
  and\ \bibinfo {author} {\bibfnamefont {N.}~\bibnamefont {Rougemaille}},\
  }\href {\doibase 10.1038/nature20155} {\bibfield  {journal} {\bibinfo
  {journal} {Nature}\ }\textbf {\bibinfo {volume} {540}},\ \bibinfo {pages}
  {410} (\bibinfo {year} {2016})}\BibitemShut {NoStop}%
\bibitem [{\citenamefont {King}\ \emph {et~al.}(2021)\citenamefont {King},
  \citenamefont {Nisoli}, \citenamefont {Dahl}, \citenamefont
  {Poulin-Lamarre},\ and\ \citenamefont {Lopez-Bezanilla}}]{King2021}%
  \BibitemOpen
  \bibfield  {author} {\bibinfo {author} {\bibfnamefont {A.~D.}\ \bibnamefont
  {King}}, \bibinfo {author} {\bibfnamefont {C.}~\bibnamefont {Nisoli}},
  \bibinfo {author} {\bibfnamefont {E.~D.}\ \bibnamefont {Dahl}}, \bibinfo
  {author} {\bibfnamefont {G.}~\bibnamefont {Poulin-Lamarre}}, \ and\ \bibinfo
  {author} {\bibfnamefont {A.}~\bibnamefont {Lopez-Bezanilla}},\ }\href
  {\doibase 10.1126/science.abe2824} {\bibfield  {journal} {\bibinfo  {journal}
  {Science}\ }\textbf {\bibinfo {volume} {373}},\ \bibinfo {pages} {576}
  (\bibinfo {year} {2021})}\BibitemShut {NoStop}%
\bibitem [{\citenamefont {Algara-Siller}\ \emph {et~al.}(2015)\citenamefont
  {Algara-Siller}, \citenamefont {Lehtinen}, \citenamefont {Wang},
  \citenamefont {Nair}, \citenamefont {Kaiser}, \citenamefont {Wu},
  \citenamefont {Geim},\ and\ \citenamefont {Grigorieva}}]{Algara2015}%
  \BibitemOpen
  \bibfield  {author} {\bibinfo {author} {\bibfnamefont {G.}~\bibnamefont
  {Algara-Siller}}, \bibinfo {author} {\bibfnamefont {O.}~\bibnamefont
  {Lehtinen}}, \bibinfo {author} {\bibfnamefont {F.~C.}\ \bibnamefont {Wang}},
  \bibinfo {author} {\bibfnamefont {R.~R.}\ \bibnamefont {Nair}}, \bibinfo
  {author} {\bibfnamefont {U.}~\bibnamefont {Kaiser}}, \bibinfo {author}
  {\bibfnamefont {H.~A.}\ \bibnamefont {Wu}}, \bibinfo {author} {\bibfnamefont
  {A.~K.}\ \bibnamefont {Geim}}, \ and\ \bibinfo {author} {\bibfnamefont
  {I.~V.}\ \bibnamefont {Grigorieva}},\ }\href {\doibase 10.1038/nature14295}
  {\bibfield  {journal} {\bibinfo  {journal} {Nature}\ }\textbf {\bibinfo
  {volume} {519}},\ \bibinfo {pages} {443} (\bibinfo {year}
  {2015})}\BibitemShut {NoStop}%
\bibitem [{\citenamefont {Benton}\ \emph {et~al.}(2016)\citenamefont {Benton},
  \citenamefont {Sikora},\ and\ \citenamefont {Shannon}}]{Benton2016}%
  \BibitemOpen
  \bibfield  {author} {\bibinfo {author} {\bibfnamefont {O.}~\bibnamefont
  {Benton}}, \bibinfo {author} {\bibfnamefont {O.}~\bibnamefont {Sikora}}, \
  and\ \bibinfo {author} {\bibfnamefont {N.}~\bibnamefont {Shannon}},\ }\href
  {\doibase 10.1103/PhysRevB.93.125143} {\bibfield  {journal} {\bibinfo
  {journal} {Phys. Rev. B}\ }\textbf {\bibinfo {volume} {93}},\ \bibinfo
  {pages} {125143} (\bibinfo {year} {2016})}\BibitemShut {NoStop}%
\bibitem [{\citenamefont {Kond\'akor}\ and\ \citenamefont
  {Penc}(2023)}]{Kondakor2023}%
  \BibitemOpen
  \bibfield  {author} {\bibinfo {author} {\bibfnamefont {M.}~\bibnamefont
  {Kond\'akor}}\ and\ \bibinfo {author} {\bibfnamefont {K.}~\bibnamefont
  {Penc}},\ }\href {\doibase 10.1103/PhysRevResearch.5.043172} {\bibfield
  {journal} {\bibinfo  {journal} {Phys. Rev. Res.}\ }\textbf {\bibinfo {volume}
  {5}},\ \bibinfo {pages} {043172} (\bibinfo {year} {2023})}\BibitemShut
  {NoStop}%
\bibitem [{\citenamefont {Nagle}(1966)}]{Nagle1966}%
  \BibitemOpen
  \bibfield  {author} {\bibinfo {author} {\bibfnamefont {J.~F.}\ \bibnamefont
  {Nagle}},\ }\href {\doibase 10.1063/1.1705058} {\bibfield  {journal}
  {\bibinfo  {journal} {Journal of Mathematical Physics}\ }\textbf {\bibinfo
  {volume} {7}},\ \bibinfo {pages} {1484} (\bibinfo {year} {1966})}\BibitemShut
  {NoStop}%
\bibitem [{\citenamefont {Schecter}\ and\ \citenamefont
  {Iadecola}(2018)}]{Iadecola2018}%
  \BibitemOpen
  \bibfield  {author} {\bibinfo {author} {\bibfnamefont {M.}~\bibnamefont
  {Schecter}}\ and\ \bibinfo {author} {\bibfnamefont {T.}~\bibnamefont
  {Iadecola}},\ }\href {\doibase 10.1103/PhysRevB.98.035139} {\bibfield
  {journal} {\bibinfo  {journal} {Phys. Rev. B}\ }\textbf {\bibinfo {volume}
  {98}},\ \bibinfo {pages} {035139} (\bibinfo {year} {2018})}\BibitemShut
  {NoStop}%
\end{thebibliography}%
\end{document}